\documentclass{ws-ijmpe}
\usepackage[super,compress]{cite}
\begin{document}

\markboth{L.A. Malov et al.}{Excitation energies of  $2^+_1$ and $4^+_1$ states of neutron deficient U and Pu isotopes}

\catchline{}{}{}{}{}

\title{Excitation energies of  $2^+_1$ and $4^+_1$ states of neutron deficient U and Pu isotopes}

\author{L.A. Malov, N.Yu. Shirikova}
\address{Bogoliubov Laboratory of Nuclear Physics\\
Joint Institute for Nuclear Research, 141980 Dubna, Russia}

\author{R. V. Jolos\footnote{Corresponding author}, E. A. Kolganova}
\address{Bogoliubov Laboratory of Nuclear Physics\\
Joint Institute for Nuclear Research, 141980 Dubna, Russia,\\
Dubna State University, 141982 Dubna, Russia\\
*jolos@theor.jinr.ru}

\maketitle

\begin{history}
\end{history} 

\begin{abstract}

The microscopic variant of the Grodzins relation and the Quasiparticle Phonon Model are applied to predict the excitation energies of the $2^+_1$  states of neutron deficient U and Pu isotopes. The P-factor systematics is used to determine the quadrupole deformation of nuclei under consideration. The excitation energies of the $4^+_1$ states are predicted based on the simple universal anharmonic vibrator type relation.

\end{abstract} 


\ccode{PACS numbers: 21.10.Re, 21.10.Dr}


\section{Introduction}

The study of the structure of nuclei belonging to the isotopic chains of various elements, including both spherical and deformed nuclei,  allows one to obtain information about the change in the average field and the structure of  excited states during the evolution of the shape of  nuclei. Such isotopic chains are known, for example, in the region of  rare earth elements. It would be interesting to obtain  information on long isotopic chains also in the region of actinides and transfermium elements. In these nuclei, neutrons fill the shell N$=126 - 184$ and it would be interesting to get information about the properties of nuclei at the beginning of this shell.  Synthesis and experimental studies of these nuclei are carried out at the Flerov Laboratory of Nuclear Reactions in Dubna~\cite{Eremin1,Eremin2,Eremin3}.

It is well known that an important indicator of the shape and other properties of even-even nuclei is the excitation energy of their first excited $2^+$ ($E(2^+_1)$) and $4^+$ ($E(4^+_1)$) states. For this reason, information on $E(2^+_1)$ and 
$E(4^+_1)$ energies of even-even nuclei becomes very important for understanding  their structure. For experiments planned to measure $E(2^+_1)$ it could  also be useful to know the theoretical predictions of  $E(2^+_1)$ values.

\section{Methods and Results}

The well-known  phenomenological relation for the product of the energy of the $2^+_1$ state per probability of the E2 transition from the ground state to the $2^+_1$ state was  established by Grodzins  in 1962 \cite{Grodzins}. In Ref.\cite{Jolos1}, the phenomenological Grodzins relation was 
 derived theoretically based on the collective quadrupole Bohr Hamiltonian. As a result, the following relation for $E(2^+_1)$ was obtained:
\begin{eqnarray}
\label{Eq13}
E(2^+_1)=\hbar^2\frac{1}{\beta^2_2}\left(\frac{2}{5}\frac{1}{B_{rot}}+\frac{2}{5}\frac{1}{B_{\gamma}}+\frac{1}{5}\frac{1}{B_{\beta}}\right),
\end{eqnarray}
where $\beta_2$ is the quadrupole deformation and $B_{rot}$, $B_{\gamma}$ and $B_{\beta}$ are the inertia coefficients for rotational, $\gamma$- and $\beta$-motions, respectively.
The expressions for  the inertia coefficients have been derived in the framework of the microscopical nuclear model  using the cranking approximation:
\begin{eqnarray}
\label{Eq14}
&\!\!B_{rot}&\!\!=2\hbar^2 \sum_{s,t}\frac{|\langle s|\frac{dV}{dr}\frac{1}{\sqrt{2}}(Y_{21}+Y_{2-1})|t\rangle|^2\left(\varepsilon_s\varepsilon_t-(E_s-\lambda)(E_t-\lambda)-\Delta_s\Delta_t\right)}{2\varepsilon_s\varepsilon_t(\varepsilon_s+\varepsilon_t)^3}, \nonumber \\ 
\label{Eq15}
&\!\!B_{\beta}&\!\!=2\hbar^2 \sum_{s,t}\frac{|\langle s|\frac{dV}{dr}Y_{20}|t\rangle|^2\left(\varepsilon_s\varepsilon_t-(E_s-\lambda)(E_t-\lambda)+\Delta_s\Delta_t\right)(\varepsilon_s+\varepsilon_t)}{2\varepsilon_s\varepsilon_t((\varepsilon_s+\varepsilon_t)^2-\omega_{\beta}^2)^2},\\
\label{Eq16}
&\!\!B_{\gamma}=&\!2\hbar^2 \sum_{s,t}\frac{|\langle s|\frac{dV}{dr}\frac{1}{\sqrt{2}}(Y_{22}+Y_{2-2})|t\rangle|^2\left(\varepsilon_s\varepsilon_t-(E_s-\lambda)(E_t-\lambda)+\Delta_s\Delta_t\right)(\varepsilon_s+\varepsilon_t)}{2\varepsilon_s\varepsilon_t((\varepsilon_s+\varepsilon_t)^2-\omega_{\gamma}^2)^2}. \nonumber
\end{eqnarray}
In these relations $V$ is the Woods-Saxon nuclear mean field potential,
$Y_{2\mu}$ is the spherical function, $(E_s-\lambda)$ is the single particle
energy, $\varepsilon$ is the single quasiparticle energy, $\Delta_s$ is the
nuclear pairing energy parameter corresponding to the single particle
state with quantum number $s$, and $\omega_{\beta}(\omega_{gamma})$ is the
energy of the $\beta (\gamma)$ vibrational phonon.
The quantities presented in (\ref{Eq15}) 
are calculated below for the neutron deficient U and Pu isotopes using the Quasiparticle Phonon Model \cite{Soloviev1,Soloviev2}.  The details of calculations are given in Ref.\cite{Shirikova}.

In the case of deformed nuclei, the experimental information on the ratios $B_{\gamma}$/$B_{rot}$ and $B_{\beta}$/$B_{rot}$
was analyzed in Refs.\cite{Jolos2,Jolos3,Jolos4}. It was shown that the ratio $B_{\gamma}$/$B_{rot}$ fluctuates around 4.2 and
 $B_{\beta}$/$B_{rot}$ fluctuates around 11.8. Based on this information we fix $B_{\gamma}$/$B_{rot}=4$ and
$B_{\beta}$/$B_{rot}=12$.  Then instead of (\ref{Eq13}) we obtain the following relation:
\begin{equation}
\label{EqA}
E(2^+_1)=\frac{0.52\hbar^2}{\beta^2_2}B_{rot}.
\end{equation}
Both expressions for $E(2^+_1)$ (\ref{Eq13}) and (\ref{EqA}) are used below to calculate the energies of the $2^+_1$ states
of the neutron deficient U and Pu isotopes.

The values of $\beta_2$ required for calculations are determined as follows. In the case of nuclei for which experimental information is available, experimental values are used in the calculations. If there is no experimental information,
$\beta_2$ values are determined in the following way. It is known that nuclei  for which the values of the $P$-factor coincide are characterized by close values of the characteristics of low-lying collective quadrupole states \cite{Casten1}. The $P$-factor is determined by the expression
\begin{equation}
\label{Cas1}
P=\frac{N_p N_n}{N_p+N_n},
\end{equation}
where $N_p$ ($N_n$) is the number of valence protons (neutrons). For those nuclei for which there is no experimental information about $\beta_2$, we find a nucleus with a close value of the $P$-factor and an experimentally known value of $\beta_2$.
This value of $\beta_2$  is taken as the value of the quadrupole deformation for the nucleus under consideration. For instance, for $^{222}$U ($P =2.86$),  $^{222}$Ra ($P =2.86$) is taken,  for $^{228}$U ($P =5.00$),  $^{230}$Th ($P =5.09$), is taken, and so on.
\begin{table*}[hbt]
\centering
\caption{Quadrupole deformation, and the calculated and experimental excitation energies of the $2^+_1$ and $4^+_1$  states of the neutron deficient U and Pu isotopes.  Calculations are based on  equations (\ref{Eq13}) (third column) and (\ref{EqA}) (fourth column). Energies are given in keV.}
\label{table:one}
\begin{tabular}{l||c|c|c|c|c|c}
\hline
Nucleus  & $\beta_{20}$ &  $E(2^+_1)_{cal}$    &  $E(2^+_1)_{cal}$   & $E(2^+_1)_{exp}$  & $E(4^+_1)_{cal}$  & $E(4^+_1)_{exp}$\\  
         &              & Eq.(\ref{Eq13})        &  Eq.(\ref{EqA})       &                   &                   &   \\
\hline
$^{222}$U & 0.142  & 231  & 258  & -- & 542-596 & --  \\
$^{224}$U & 0.179  & 155  & 156  & -- & 390-392 & --  \\
$^{226}$U & 0.228  &  84  &  87  & 81 & 248-254 & 250   \\
$^{228}$U & 0.245  &  77  &  80  & 59 & 234-240 & -- \\
$^{226}$Pu & 0.202 &  88  &  95  & -- & 256-270 & --  \\
$^{228}$Pu & 0.230 &  64  &  68  & -- & 208-216 & --  \\
$^{230}$Pu & 0.261 &  51  &  53  & -- & 182-186 & -- \\
$^{232}$Pu & 0.272 &  45  &  46  & -- & 170-172 & --\\

\hline
\end{tabular}
\end{table*}

The results obtained using relations (\ref{Eq13}) and (\ref{EqA}) are shown in Table~\ref{table:one}. In order to have an estimate of the error made on the basis of equations (\ref{Eq13}) and (\ref{EqA}), Table~\ref{table:two} shows the calculated energies of the $2^+_1$ states of the well studied $^{230-238}$U together with experimental values. It is seen from Tables~\ref{table:one} and \ref{table:two} that the energies of the $2^+_1$ states mostly increase with decreasing mass number following a decrease in the quadrupole deformation. However, as it is seen from Eq.(\ref{Eq13}) the energy $E(2^+_1)$ depends not only on the quadrupole deformation $\beta_2$ but also on the inertia coefficients $B_{rot}$, $B_{\gamma}$ and $B_{\beta}$. As it is seen from Eq. (\ref{Eq14}) some tiny details in the single particle level scheme can be a reason of a small decrease in the excitation energy of the $2^+_1$ state inspite a decrease of the quadrupole deformation.   

In addition to $E(2^+_1)$ we have also calculated the values of $E(4^+_1)$. It was discovered in \cite{Casten2} as a result of an analysis of experimental data that the excitation energies of the $E(4^+_1)$ states of even-even nuclei with
$2\le E(4^+_1)/E(2^+_1)\le 3.14$ satisfy to the universal relation
\begin{equation}
\label{Cas2}
E(4^+_1)=2E(2^+_1)+\epsilon_4,
\end{equation}
where the parameter $\epsilon_4$ is a constant for a given region of the nuclide chart. For the neutron deficient U and Pu isotopes
$\epsilon_4$ takes the values around $67 \pm 16$ keV. This relation is similar to the anharmonic vibrator equation with constant anharmonicity $\epsilon_4$. Later, this relation was derived  in the framework of the Interacting Boson Model \cite{Jolos5}.

Considering experimental data for nuclei with the $P$-factor close to that for the neutron deficient U and Pu isotopes, we take 
$\epsilon_4=80$ keV. Using this value and the relation (\ref{Cas2}), we have calculated the excitation energies of the 
$4^+_1$ states. The results of calculations are presented in Table~\ref{table:one}.

In Table~\ref{table:three},  the excitation energies and the $P$-factors for $^{222-228}$U, $^{226-232}$Pu  and those nuclei whose values of the $P$-factor are close to the values of the $P$-factor for one of the $^{222-228}$U or $^{226-232}$Pu isotopes   are given. For these nuclei the experimental values of $E(2^+_1)$ are known. Thus, comparing the values of $E(2^+_1)$ given in columns 3 and 6 of Table III, we obtain additional information on the accuracy of the predictions made on the basis of Eqs.(\ref{Eq13}) and (\ref{EqA}) or about the predictive capabilities of the $P$-systematic scheme.


\begin{table*}[hbt]
\centering
\caption{Quadrupole deformation, and the calculated and experimental excitation energies of the $2^+_1$ states of the well studied U isotopes.  Calculations are based on  equations (\ref{Eq13}) (third column) and (\ref{EqA}) (fourth column). Energies are given in keV.}
\label{table:two}
\begin{tabular}{l||c|c|c|c}
\hline
Nucleus  & $\beta_{20}$ &  $E(2^+_1)_{cal}$    &  $E(2^+_1)_{cal}$   & $E(2^+_1)_{exp}$   \\  
         &              &  Eq.(\ref{Eq13})        &  Eq.(\ref{EqA})       &                     \\
\hline
$^{230}$U  & 0.262 (exp) &  58 & 59 & 52  \\
$^{232}$U  & 0.264 (exp) &  55 & 56 & 48  \\
$^{234}$U  & 0.272 (exp) &  45 & 48 & 43  \\
$^{236}$U  & 0.282 (exp) &  39 & 43 & 45  \\
$^{238}$U  & 0.286 (exp) &  30 & 38 & 45  \\

\hline
\end{tabular}
\end{table*}


\begin{table*}[hbt]
\centering
\caption{$P$-factor and excitation energies of the $2^+_1$ states. The left half of the table shows the results for the neutron deficient U and Pu isotopes. Where the range of possible excitation energy values  is indicated, the boundaries of the range are the energies obtained using Eq.(\ref{Eq13}) and Eq.(\ref{EqA}). The right half of the table shows  data for nuclei having
$P$-factor values close to those of one of the $^{222-228}$U and $^{226-232}$Pu isotopes.}
\label{table:three}
\begin{tabular}{l|c|c||l|c|c}
\hline
Nucleus  & $P=\frac{N_pN_n}{N_p+N_n}$ &  $E(2^+_1)_{cal}$ (keV) & Nucleus & $P=\frac{N_pN_n}{N_p+N_n}$ & $E(2^+_1)_{exp}$ (keV)  \\  
\hline
$^{222}$U & 2.86  & 231-258  & $^{222}$Ra  & 2.86 & 111  \\
$^{224}$U & 3.75  & 155-156  & $^{224}$Ra  & 3.75 & 84   \\
$^{226}$U & 4.44  &  84-87   & $^{226}$Th  & 4.44 & 72    \\
$^{228}$U & 5.00  &  77-80   & $^{230}$Th  & 5.09 & 53  \\
$^{226}$Pu & 4.00 &  88-95   & $^{226}$Ra  & 4.00 & 68   \\
$^{228}$Pu & 4.80 &  64-68   & $^{228}$Th  & 4.80 & 58   \\
$^{230}$Pu & 5.45 &  51-52   & $^{232}$Th  & 5.33 & 49  \\
$^{232}$Pu & 6.00 &  45      & $^{234}$U   & 6.15 & 43 \\
\hline
\end{tabular}
\end{table*}

\section{Conclusion}

Based on the microscopic variant of the Grodzins relation and the Quasiparticle Phonon Model of the nuclear structure,
the excitation energies of the first $2^+$ states of neutron deficient even-even U and Pu isotopes are predicted. 
The excitation energies of the $4^+_1$ states are also predicted for these nuclei based on a simple universal
anharmonic vibrator type relation. The systematics of the properties of  low-lying collective states of  even-even nuclei based on their $P$-factor dependence is used to estimate  errors in the predictions.


\end{document}